\documentclass{article}
\usepackage{amsfonts}

\begin{document}

\title{Information capacity of quantum observable}
\author{A. S. Holevo}
\date{}
\maketitle

\begin{abstract}
In this paper we consider the classical capacities of quantum-classical
channels corresponding to measurement of observables. Special attention is
paid to the case of continuous observables. We give the formulas for
unassisted and entanglement-assisted classical capacities $C,C_{ea}$ and
consider some explicitly solvable cases which give simple examples of
entanglement-breaking channels with $C<C_{ea}.$
We also elaborate on the ensemble-observable duality to show that $C_{ea}$ for the measurement channel is related to the $\chi$-quantity for the
dual ensemble in the same way as $C$ is related to the accessible information. This provides both accessible information and the $\chi$-quantity for the
quantum ensembles dual to our examples.
\end{abstract}

\section{\protect\bigskip Introduction}

In quantum information theory one often has to deal with both quantum and
classical information. A usual device is then to embed the classical system
into quantum by representing classical states, i.e. probability
distributions on the phase space $\Omega ,$ as diagonal density operators in
the artificial Hilbert space $\mathcal{H}$ spanned by the orthonormal basis $
\left\{ |\omega \rangle ;\omega \in \Omega \right\} :$
\[
P=\left\{ p_{\omega }\right\} \longrightarrow \rho =\sum\limits_{\omega
}p_{\omega }|\omega \rangle \langle \omega |.
\]
This works for finite and countable $\Omega $, although in the last case $
\mathcal{H}$ becomes infinite dimensional. Any channel with discrete
classical input alphabet $\mathcal{X}$ or output alphabet $\mathcal{Y}$ can
then be regarded as a quantum channel. In the case of classical-quantum
(c-q) channel corresponding to preparation of states $\left\{ \rho _{x};x\in
\mathcal{X}\right\} $ the quantum channel is
\begin{equation}
\mathcal{P}(\rho )=\sum\limits_{x\in \mathcal{X}}\langle x|\rho |x\rangle
\rho _{x},  \label{prep}
\end{equation}
where $\left\{ |x\rangle ;x\in \mathcal{X}\right\} $ is a fixed orthonormal
basis. Similarly, in the case of quantum-classical (q-c) channel
corresponding to measurement of observable given by discrete probability
operator-valued measure (POVM) $M=\{M_{y};y\in \mathcal{Y}\}$ \cite{asp} we
have
\begin{equation}
\mathcal{M}(\rho )=\sum\limits_{y\in \mathcal{Y}}\left( \mathrm{Tr}\rho
M_{y}\right) |y\rangle \langle y|.  \label{meas}
\end{equation}

\bigskip However, in the case of continuous classical variables the
situation is different for c-q and q-c channels. In principle, there is no
problem with embedding c-q channels. Let $\mathcal{X}$ be a domain in $
\mathbb{R}^{k}$ and $dx$ is the Lebesgue measure, then the continuous analog
of (\ref{prep}) is
\[
\mathcal{P}(\rho )=\int_{\mathcal{X}}\langle x|\rho |x\rangle \rho _{x}dx,
\]
where $\left\{ |x\rangle ;x\in \mathcal{X}\right\} $ is the Dirac's system
satisfying $\langle x|x^{\prime }\rangle =\delta (x-x^{\prime }).$ Here $
\mathcal{P}$ maps density operators into density operators. Now let $
M=\{M(dy)\}$ be a quantum observable (POVM)\ with continuous set of outcomes
$\mathcal{Y}\subseteq \mathbb{R}^{k}.$ Then for a density operator $\rho $
the \textquotedblleft diagonal\textquotedblright\ operator $\int_{\mathcal{Y}
}|y\rangle \langle y|\mathrm{Tr}\rho M(dy)$ has infinite trace, so in
general there is no continuous analog of (\ref{meas}). This is related to
the well known repeatability issue for continuous observables and
nonexistence of the normal expectation onto the Abelian subalgebra of the
diagonal operators (see e.g. \cite{dav}, Sec. 4.4, \cite{ozawa}, \cite{struc}, Sec. 4.1.4).

The only way is to consider the q-c channel as transformation $\mathcal{M}
:\rho \longrightarrow \mathrm{Tr}\rho M(dy)$ of density operators to
probability distributions\footnote{
For a rigorous unified description of quantum and classical systems using
the language of operator algebras see e.g. \cite{struc}, \cite{bl}.} on $
\mathcal{Y}$. The main interest in this paper will be the classical
capacities of such channel -- both unassisted $C\left( \mathcal{M}\right) $
and entanglement-assisted $C_{ea}\left( \mathcal{M}\right)$. We give the
general formulas and consider some explicitly solvable cases which provide simple
examples of entanglement-breaking channels with $C_{ea}>C.$

Almost simultaneously with the first version of this work the papers \cite{da}, \cite{or} appeared, where the
quantity $C(\mathcal{M})$ was studied in detail for the finite case. In particular, the ensemble-observable duality
\cite{hall} was used to relate $C(\mathcal{M})$ with the accessible information of the dual ensemble.  In the last Section we elaborated further
on the duality transformation to show that $C_{ea}\left( \mathcal{M}\right)$ is in similar relation with the $\chi$-quantity for the
dual ensemble. This allows to compute both accessible information and the $\chi$-quantity for the
quantum ensembles dual to our examples.

\section{The classical capacities of quantum observables}

Consider the channel (\ref{meas}) in the case of discrete $\mathcal{Y}$ and
finite-dimensional input Hilbert space $\mathcal{H}.$ Since q-c channel is
entanglement-breaking, the unassisted classical capacity is given by the
one-letter expression
\begin{equation}
C(\mathcal{M})=C_{\chi }(\mathcal{M})=\sup_{\pi }I\left( \pi ;M\right),  \label{C}
\end{equation}
where $\pi $ is a finite probability distribution  on the
state space $\mathfrak{S}(\mathcal{H})$ assigning probabilities $\pi _{x}$
to states $\rho _{x}$ (ensemble), and
\[
I\left( \pi ;M\right) =H\left( P_{\bar{\rho}_{\pi }}\right)
-\sum\limits_{x}\pi _{x}H\left( P_{\rho _{x}}\right)
\]
is the Shannon information between the input $x$ and the output $y.$ Here $
\bar{\rho}_{\pi }=\sum\limits_{x}\pi _{x}\rho _{x}$, $P_{\rho }=\left\{
\mathrm{Tr}\rho M_{y}\right\} $ -- the probability distribution of the
measurement outcomes and $H\left( \cdot \right) $ is the Shannon entropy.
This can be rewritten as
\begin{equation}
C(\mathcal{M})=C_{\chi }(\mathcal{M})=\sup_{\rho }\chi _{\Phi }\left( \rho
\right) ,  \label{Cchi}
\end{equation}
where
\begin{equation}
\chi _{\Phi }\left( \rho \right) =H\left( P_{\rho }\right) -\inf_{\pi :\bar{
\rho}_{\pi }=\rho }\sum\limits_{x}\pi _{x}H\left( P_{\rho _{x}}\right) ,
\label{chi}
\end{equation}

Then consider the entanglement-assisted capacity which according to the
result of Shor et al \cite{bsst} is given by the formula

\begin{equation}
C_{ea}(\mathcal{M})=\sup_{\rho }I(\rho ;\mathcal{M}),  \label{Cea}
\end{equation}
where
\[
I(\rho ;\mathcal{M})=S(\rho )+S(\mathcal{M}(\rho ))-S(\rho ,\mathcal{M})
\]
is the quantum mutual information. Here $S(\cdot )$ is von Neumann
entropy and $S(\rho ,\mathcal{M})$ is the entropy exchange. Let $p_{y}=
\mathrm{Tr}\rho M_{y}$ and $V_{y}$ be an operator satisfying $
M_{y}=V_{y}^{\ast }V_{y},$ for example, $V_{y}=M_{y}^{1/2}.$ Then the
density operator $\frac{V_{y}\rho V_{y}^{\ast }}{p_{y}}=\rho \left(
y|M\right) $ can be interpreted as posterior state of the measurement of
observable $M$ with the instrument $\rho \rightarrow \{V_{y}\rho V_{y}^{\ast
}\}$ in the state $\rho $. The following formula was obtained by Shirokov
\cite{shir}
\begin{equation}
I(\rho ;\mathcal{M})=S\left( \rho \right) -\sum\limits_{y}\left( \mathrm{Tr}
\rho M_{y}\right) S\left( \rho \left( y|M\right) \right) .  \label{I}
\end{equation}
Indeed, let us use the relation $S\left( \rho ,\mathcal{M}\right) =S(\tilde{
\mathcal{M}}(\rho )),$ where $\tilde{\mathcal{M}}$ is the complementary
channel. According to \cite{comp} the complementary channel
for (\ref{meas}) is
\[
\tilde{\mathcal{M}}(\rho )=\sum\limits_{y}|y\rangle \langle y|\otimes
V_{y}\rho V_{y}^{\ast }=\sum\limits_{y}|y\rangle \langle y|\otimes p_{y}\rho
\left( y|M\right) .
\]
It follows $S(\tilde{\mathcal{M}}(\rho ))=H(P_{\rho
})+\sum\limits_{y}p_{y}S\left( \rho \left( y|M\right) \right) ,$ while $S(
\mathcal{M}(\rho ))=H(P_{\rho })$, hence (\ref{I}). The
entanglement-assisted classical capacity of the channel $\mathcal{M}$
follows by substituting this expression into (\ref{Cea}).

Now consider the channel $\mathcal{M}:\rho \longrightarrow \mathrm{Tr}\rho
M(dy)$ in the case of arbitrary measurable space $\mathcal{Y}$ and
finite-dimensional input Hilbert space $\mathcal{H}.$ Then the relations
(\ref{C}) and (\ref{Cea}) can be generalized to this case. Since the output
is classical, the protocol of entanglement-assisted transmission of
classical information should be explained in this case. First, a pure
entangled state
\[
|\psi \rangle =\sum_{j}\lambda _{j}|j\rangle \otimes |j\rangle ,
\]
where $\left\{ |j\rangle \right\} $ is an orthonormal basis in $\mathcal{H}$
is distributed between the input (Alice) and output (Bob). Thus classical
Bob gets additional quantum space $\mathcal{H}$ becoming classical-quantum
system \cite{bl}. The states of such a system are positive operator-valued
measures $\left\{ \sigma (dy)\right\} $ satisfying $\mathrm{Tr}\int_{
\mathcal{Y}}\sigma (dy)=1.$ Alice uses different encoding maps $\mathcal{E}
_{w}$ for different input signals $w$. The joint state of Alice and Bob is
then
\[
\left( \mathcal{E}_{w}\otimes \mathrm{Id}\right) \left( |\psi \rangle
\langle \psi |\right) =\sum_{j,k}\lambda _{j}\lambda _{k}\mathcal{E}
_{w}\left( |j\rangle \langle k|\right) \otimes |j\rangle \langle k|,
\]
and after the Alice measurement $\mathcal{M}$ Bob gets the state $\left\{
\sigma _{w}(dy)\right\} $ with
\[
\sigma _{w}(dy)=\sum_{j,k}\lambda _{j}\lambda _{k}\left[ \mathrm{Tr}\mathcal{
E}_{w}\left( |j\rangle \langle k|\right) M(dy)\right] |j\rangle \langle k|
\]
\[
=\sum_{j,k}\lambda _{j}\lambda _{k}\langle k|\mathcal{E}_{w}^{\ast }\left(
M(dy)\right) |j\rangle |j\rangle \langle k|=\rho ^{1/2}\overline{\mathcal{E}
_{w}^{\ast }\left( M(dy)\right) }\rho ^{1/2},
\]
where $\rho =\sum_{j}\lambda _{j}|j\rangle \langle j|$ and bar means complex
conjugation in the basis $\left\{ |j\rangle \right\} .$ Then Bob applies his
decoding given by classical-quantum observable $\left\{ N_{yw^{\prime
}}\right\} ,$ satisfying $\sum_{w}N_{yw^{\prime }}\equiv I,$ with the
conditional probabilities of outcomes $\mathsf{P}\left( w^{\prime }|w\right)
=\int_{\mathcal{Y}}\mathrm{Tr}\sigma _{w}(dy)N_{yw^{\prime }}.$

The continuous analog of formula (\ref{I}) considered in \cite{shir} is
\begin{equation}
I(\rho ;\mathcal{M})=S\left( \rho \right) -\int_{\mathcal{Y}}\left( \mathrm{
Tr}\rho M(dy)\right) S\left( \rho \left( y|M\right) \right) .  \label{I_mod}
\end{equation}

In \cite{bsst} it was stressed that entanglement-assisted classical capacity
of entanglement-breaking channels can be greater than the unassisted
capacity. The example given there was the depolarizing channel with high
enough error probability (see also Appendix). In the next Sections we will see that the
inequality $C_{ea}>C$ is rather common for the measurement channels with unsharp observables.

\section{Examples}

\textbf{1. }Consider the case of general overcomplete system, $M_{y}=|\psi
_{y}\rangle \langle \psi _{y}|,$ $\sum_{y}|\psi _{y}\rangle \langle \psi
_{y}|=I$ in $m-$dimensional Hilbert space $\mathcal{H}.$ Then the posterior
state $\rho \left( y|M\right) =\frac{|\psi _{y}\rangle \langle \psi _{y}|}{
\langle \psi _{y}|\psi _{y}\rangle }$ is pure and hence $S\left( \rho \left(
y|M\right) \right) =0.$ Thus $I(\rho ;\mathcal{M})=S\left( \rho \right) $ and
\begin{equation}
C_{ea}(\mathcal{M})=\sup_{\rho }S\left( \rho \right) =\log m.  \label{C1}
\end{equation}

A special case is covariant observable
\begin{equation}
M_{g}=\frac{m}{\left\vert G\right\vert }V_{g}|\psi _{0}\rangle \langle \psi
_{0}|V_{g}^{\ast },  \label{cov}
\end{equation}
where $V_{g}$ is irreducible representation of the group $G$ and $|\psi
_{0}\rangle $ is a unit vector \cite{asp}. Then the channel $\mathcal{M}$ is
covariant and by \cite{covar} we have
\begin{equation}
C(\mathcal{M})=C_{\chi }(\mathcal{M})=H\left( \mathcal{M}\left( \frac{I}{m}
\right) \right) -\min_{\psi }H\left( \mathcal{M}\left( |\psi \rangle \langle
\psi |\right) \right) .  \label{C3}
\end{equation}
But $\mathcal{M}\left( \frac{I}{m}\right) $ is uniform distribution over $G,$
hence $H\left( \mathcal{M}\left( \frac{I}{m}\right) \right) =\log \left\vert
G\right\vert ,$ while
\begin{eqnarray*}
H\left( \mathcal{M}\left( |\psi \rangle \langle \psi |\right) \right)
&=&-\sum_{g}\frac{m}{\left\vert G\right\vert }\left\vert \langle \psi
|V_{g}|\psi _{0}\rangle \right\vert ^{2}\log \frac{m}{\left\vert
G\right\vert }\left\vert \langle \psi |V_{g}|\psi _{0}\rangle \right\vert
^{2} \\
&=&-\log \left\vert G\right\vert -\frac{m}{\left\vert G\right\vert }
\sum_{g}\left\vert \langle \psi |V_{g}|\psi _{0}\rangle \right\vert ^{2}\log
\left\vert \langle \psi |V_{g}|\psi _{0}\rangle \right\vert ^{2}.
\end{eqnarray*}
Therefore
\[
C(\mathcal{M})=\log m+\frac{m}{\left\vert G\right\vert }\max_{\psi
}\sum_{g}\left\vert \langle \psi |V_{g}|\psi _{0}\rangle \right\vert
^{2}\log \left\vert \langle \psi |V_{g}|\psi _{0}\rangle \right\vert ^{2}
\]
which is typically less than $C_{ea}(\mathcal{M})=\log m$ (see Sec. 4).

\bigskip

\textbf{2.} Let $\Theta $ be the unit sphere in $\mathcal{H}$ and let $\nu
(d\theta )$ be the uniform distribution on $\Theta .$ Then by \cite{asp} ,
Sec. IV.4 (see also Appendix)
\begin{equation}
m\int_{\Theta }|\theta \rangle \langle \theta |\nu (d\theta )=I,
\label{resid}
\end{equation}
thus we have a continuous overcomplete system, i.e. observable
$M(d\theta)=m|\theta \rangle \langle \theta |\nu (d\theta )$ in $\mathcal{H}$ with
values in $\Theta $ . According to the remark above, $C_{ea}(\mathcal{M})=\log m.$

The channel $\mathcal{M}$ maps density operator $\rho $ to the probability
distribution $m\langle \theta |\rho |\theta \rangle \nu (d\theta )$ on $
\Theta .$ All these outcome probability distributions are absolutely
continuous with respect to $\nu (d\theta ),$ hence we can use the
differential entropy $h(p_{\rho })=-\int_{\Theta }p_{\rho }(\theta )\log
p_{\rho }(\theta )\nu (d\theta ),$ where $p_{\rho }(\theta )=$ $m\langle
\theta |\rho |\theta \rangle ,$ to get the continuous analog of the formula
(\ref{chi})
\begin{equation}
\chi _{\Phi }\left( \rho \right) =h\left( p_{\rho }\right) -\inf_{\pi :\bar{
\rho}_{\pi }=\rho }\sum\limits_{x}\pi _{x}h\left( p_{\rho _{x}}\right) .
\label{chi_mod}
\end{equation}
Let us use this formula together with (\ref{Cchi}) to compute
$C(\mathcal{M})=C_{\chi }(\mathcal{M}).$

The channel $\mathcal{M}$ is covariant with respect to the irreducible
action of the unitary group $U(\mathcal{H})$ in the sense that
\[
\mathcal{M}\left( U\rho U^{\ast }\right) =m\langle U^{\ast }\theta |\rho
|U^{\ast }\theta \rangle \nu (d\theta ).
\]
Therefore similarly to (\ref{C3})
\begin{equation}
C(\mathcal{M})=C_{\chi }(\mathcal{M})=h\left( \mathcal{M}\left( \frac{I}{d}
\right) \right) -\min_{\theta ^{\prime }}h\left( \mathcal{M}\left( |\theta
^{\prime }\rangle \langle \theta ^{\prime }|\right) \right) ,  \label{C2}
\end{equation}
But $\mathcal{M}\left( \frac{I}{d}\right) $ is uniform distribution over $
\Theta $ with the density $p(\theta )\equiv 1,$ hence $h\left( \mathcal{M}
\left( \frac{I}{d}\right) \right) =0.$ On the other hand,
\begin{equation}
-h\left( \mathcal{M}\left( |\theta ^{\prime }\rangle \langle \theta ^{\prime
}|\right) \right) =\int_{\Theta }m\left\vert \langle \theta |\theta ^{\prime
}\rangle \right\vert ^{2}\log m\left\vert \langle \theta |\theta ^{\prime
}\rangle \right\vert ^{2}\nu (d\theta ).  \label{int}
\end{equation}
By unitary invariance of $\nu $, this quantity is the same for all $\theta
^{\prime }$ so there is no need for minimization in (\ref{C2}). To compute
it we use Lemma IV.4.1 from \cite{asp} according to which
\begin{equation}
\int_{\Theta }F\left( \left\vert \langle \theta |\theta ^{\prime }\rangle
\right\vert \right) \nu (d\theta )=-\int\limits_{0}^{1}F(r)d(1-r^{2})^{m-1}.
\label{formula}
\end{equation}
Then (\ref{int}) becomes
\[
-\int\limits_{0}^{1}mr^{2}\log
mr^{2}d(1-r^{2})^{m-1}=\int\limits_{0}^{1}m(1-u)\log m(1-u)du^{m-1},
\]
where $u=1-r^{2},$ which after integrations by parts gives (see Appendix)
\begin{equation}
C(\mathcal{M})=\log m-\log e\sum_{k=2}^{m}\frac{1}{k}.  \label{euler}
\end{equation}
For $m\rightarrow \infty $ we have $C(\mathcal{M})\rightarrow \log e\,(1-\gamma ),$
where $\gamma \approx 0.577$ is Euler's constant. At the same time, $C_{ea}(
\mathcal{M})=\log m\rightarrow \infty .$

The value (\ref{euler}) was obtained in the paper \cite{robb} as the ``subentropy''
of the chaotic state $\rho=I/m$. This is not a simple coincidence, see Sec. 4.

\bigskip

\textbf{3.} If $\mathcal{H}$ is infinite-dimensional while $\mathcal{Y}$ is
discrete, then the quantity (\ref{C}) is usually infinite but there is
additional input constraint $\left\{ \rho :\mathrm{Tr}\rho E\leq N\right\} ,$
where $E$ \ is a positive selfadjoint operator (usually, energy) and $N$ is
a constant (energy constraint). Then instead of (\ref{Cchi}) one has the
constrained classical capacity
\begin{equation}
C(\mathcal{M},N)=C_{\chi }(\mathcal{M},N)=\sup_{\rho :\mathrm{Tr}\rho E\leq
N}\chi _{\Phi }\left( \rho \right) ,  \label{Ccon}
\end{equation}
and instead of (\ref{Cea}) -- the constrained entanglement-assisted
classical capacity
\begin{equation}
C_{ea}(\mathcal{M},N)=\sup_{\rho :\mathrm{Tr}\rho E\leq N}I(\rho ;\mathcal{M}
).  \label{Ceacons}
\end{equation}
Some additional conditions are required to ensure finiteness of entropies in
(\ref{chi}) and (\ref{I}), see \cite{const}.

If $\mathcal{M}$ is continuous observable, we expect formulas (\ref{Ccon}),
(\ref{Ceacons}) to hold with appropriate modifications (\ref{chi_mod}), resp.
(\ref{I_mod}). Such is the case of the canonical observable with the energy
constraint. Consider one Bosonic mode $Q,P$ and the canonical observable
given by POVM
\begin{equation}
M(d^{2}z)=|z\rangle \langle z|\frac{d^{2}z}{\pi };\quad z\in \mathbb{C}.
\label{canobs}
\end{equation}
This observable describes approximate joint measurement of $Q,P$ \cite{asp},
Sec. VI.8 and in quantum optics is realized by heterodyning. The
corresponding q-c channel $\mathcal{M}$ takes a density operator into the
probability distribution
\[
\rho \rightarrow p_{\rho }(z)=\langle z|\rho |z\rangle \frac{d^{2}z}{\pi },
\]
which is absolutely continuous with respect to the Lebesgue measure $\frac{
d^{2}z}{\pi }$ with the probability density $\langle z|\rho |z\rangle $
equal to the Husimi function. The posterior states are the coherent states $
\rho \left( z|M\right) =|z\rangle \langle z|$ which are pure and have zero
entropy. Thus $I(\rho ;\mathcal{M})=S\left( \rho \right) .$

Denote by $\rho _{N}$ the Gaussian density operator with zero mean and the
number of quanta $\mathrm{Tr}\rho _{N}a^{\dagger }a=N.$ It maximizes the
quantum entropy under the constraint
\begin{equation}
\mathrm{Tr}\rho a^{\dagger }a\leq N,  \label{number}
\end{equation}
namely
\[
\max_{(\ref{number})}S(\rho )=S(\rho _{N})=(N+1)\log (N+1)-N\log N\equiv
g(N).
\]
The formula (\ref{Ceacons}) gives then the following expression for the
entanglement-assisted classical capacity of channel $\mathcal{M}$ with the
constraint (\ref{number})
\begin{equation}
C_{ea}(\mathcal{M};N)=g(N)=\log (N+1)+\log \left( 1+\frac{1}{N}\right) ^{N}.
\label{C4}
\end{equation}

On the other hand, the channel is covariant with respect to the irreducible
action of the Weyl (displacement) operators at the input and the shift group
of the argument $z$ at the output. The output entropy of the channel $
\mathcal{M}$ is just the classical differential entropy $h(p_{\rho })$ and
the continuous analog of (\ref{C2}) gives
\[
C(\mathcal{M};N)=C_{\chi }(\mathcal{M};N)=\max_{(\ref{number})}h(p_{\rho })-
\check{h}(\mathcal{M}),
\]
where
\[
\check{h}(\mathcal{M})=\min_{|\psi \rangle \langle \psi |}\check{h}(p_{|\psi
\rangle \langle \psi |})
\]
is the minimal output differential entropy. By the Wehrl conjecture proved
by Lieb \cite{lieb}, $\check{h}(\mathcal{M})=\log e$ and the minimum is
attained on any coherent state. On the other hand, $\max_{(\ref{number}
)}h(p_{\rho })$ is attained on $\rho _{N}$ and
\[
\max_{(\ref{number})}h(p_{\rho })=h(p_{\rho _{N}})=\log e(N+1).
\]
Indeed,
\[
\int p_{\rho }(z)|z|^{2}\frac{d^{2}z}{\pi }=\int \langle z|a^{\dagger }\rho
a|z\rangle \frac{d^{2}z}{\pi }=\mathrm{Tr}\rho aa^{\dagger }=\mathrm{Tr}\rho
a^{\dagger }a+1
\]
and the constraint (\ref{number}) implies $\int p_{\rho }(z)|z|^{2}\frac{
d^{2}z}{\pi }\leq N+1.$ But $\max h(p)$ under the last constraint is
achieved on the probability density
\[
(N+1)^{-1}\exp \left( -\frac{|z|^{2}}{N+1}\right) =p_{\rho _{N}}(z).
\]
Thus we obtain the value\footnote{This formula as well as similar result for
homodyne measurement ($Q$ or $P$) were obtained in \cite{hall} by  ``information exclusion'' argument.}
\begin{equation}
C(\mathcal{M};N)=C_{\chi }(\mathcal{M};N)=\log (N+1).  \label{C5}
\end{equation}

\section{Ensemble-measurement duality}

Let us first describe the duality between quantum observables and ensembles, see \cite{hall},  \cite{da}, \cite{or}.
If $M=\{M_{y};y\in \mathcal{Y}\}$ is a quantum observable and  $\pi
=\left\{ p_{x},\rho _{x};x\in \mathcal{X}\right\} $  an ensemble of
quantum states then
\[
p_{xy}=p_{x}\mathrm{Tr}\rho _{x}M_{y}
\]
is a probability distribution on $\mathcal{X\times Y}.$  On the other hand,
\[
p_{xy}=p_{y}^{\prime }\mathrm{Tr}\rho _{y}^{\prime }M_{x}^{\prime },
\]
where, denoting $\bar{\rho}_{\pi }=\sum\limits_{x}p_{x}\rho _{x},$ we have $
p_{y}^{\prime }\rho _{y}^{\prime }=\bar{\rho}_{\pi }{}^{1/2}M_{y}\bar{\rho}
_{\pi }{}^{1/2}$ so that $p_{y}^{\prime }=\mathrm{Tr}\bar{\rho}_{\pi }M_{y}$
and $M_{x}^{\prime }=p_{x}\bar{\rho}_{\pi }{}^{-1/2}\rho _{x}\bar{\rho}_{\pi
}{}^{-1/2}.$ Here $M^{\prime }=\left\{ M_{x}^{\prime };x\in \mathcal{X}
\right\} $ is the new observable and $\pi ^{\prime }=\left\{ p_{y}^{\prime
},\rho _{y}^{\prime };y\in \mathcal{Y}\right\} $ is the new ensemble. Therefore the Shannon information
between $x,y$ is
\[
I(\pi ,M)\mathbf{=}I(\pi ^{\prime },M^{\prime }\mathbf{).}
\]

From this it is deduced (\cite{da}, Proposition 3) that
\begin{equation}
C(\mathcal{M})\equiv \max_{\pi }I(\pi ,M)=\max_{\rho }A(\pi _{\rho }^{\prime
}\mathbf{),}  \label{A1}
\end{equation}
where $A(\pi _{\rho }^{\prime }\mathbf{)=}\max_{M^{\prime \prime }}I(\pi
_{\rho }^{\prime },M^{\prime \prime })$ is the accessible information of the
ensemble $\pi _{\rho }^{\prime }=\left\{ \mathrm{Tr}\rho M_{y},\frac{\rho
^{1/2}M_{y}\rho ^{1/2}}{\mathrm{Tr}\rho M_{y}}\right\} .$

Let us recall the well known bound \cite{h1}
\begin{equation}\label{ho}
A(\pi )\leq S\left( \sum\limits_{y}p_{y}\rho _{y}\right) -\sum\limits_{y}p_{y}S\left( \rho
_{y}\right) \equiv \chi \left( \pi \right)
\end{equation}
with the equality attained if and only if the operators $\pi_y\rho_y$ all commute.
Applying this to the dual situation we obtain
\begin{equation}\label{dualh}
A(\pi _{\rho }^{\prime })\leq S\left( \sum\limits_{y}p_{y}^{\prime
}\rho _{y}^{\prime }\right) -\sum\limits_{y}p_{y}^{\prime }S\left( \rho
_{y}^{\prime }\right) = \chi \left( \pi _{\rho }^{\prime }\right) .
\end{equation}
But $\sum\limits_{y}p_{y}^{\prime }\rho _{y}^{\prime }=\rho ,$ and $S\left(
\rho _{y}^{\prime }\right) =H\left( \frac{V_{y}\rho V_{y}^{\ast }}{
p_{y}^{\prime }}\right) ,$ where $V_{y}$ is arbitrary operator satisfying $
M_{y}=V_{y}^{\ast }V_{y}$ because operators $V_{y}\rho V_{y}^{\ast
}=V_{y}\rho ^{1/2}\rho ^{1/2}V_{y}^{\ast }$ and $\rho ^{1/2}V_{y}^{\ast
}V_{y}\rho ^{1/2}=\rho ^{1/2}M_{y}\rho ^{1/2}=p_{y}^{\prime }\rho
_{y}^{\prime }$ are unitarily equivalent via polar decomposition and hence
have the same spectrum. The density operator $\frac{V_{y}\rho V_{y}^{\ast }
}{p_{y}^{\prime }}=\rho \left( y|M\right) $ is the posterior state of the
measurement of observable $M$ with the instrument $\{V_{y}\}$ in the state $
\rho .$ Thus
\begin{equation}
\chi \left( \pi _{\rho }^{\prime }\right) =S\left( \rho \right)
-\sum\limits_{y}p_{y}^{\prime }S\left( \rho \left( y|M\right) \right)
=I(\rho ;\mathcal{M})  \label{hbound2}
\end{equation}
i.e. the $\chi$-quantity in the right side of (\ref{ho}) is dual to the quantum mutual information for the measurement channel
and hence in addition to (\ref{A1}) we have via (\ref{I})
\begin{equation}\label{A2}
C_{ea}(\mathcal{M})=\max_{\rho }\chi \left( \pi _{\rho }^{\prime }\right) .
\end{equation}

The inequality (\ref{dualh}) appears in \cite{hall}, Eq. (19), as ``measurement-dependent
dual'' to (\ref{ho}). The necessary and sufficient condition for the equality in the case of (\ref{dualh}) becomes
\begin{equation}\label{equ}
\rho^{1/2}M_y\rho M_{y'}\rho^{1/2}=\rho^{1/2}M_{y'}\rho M_y\rho^{1/2}
\end{equation}
for all $y,y'$. Therefore necessary and sufficient condition for the equality
$C_{ea}(\mathcal{M})=C(\mathcal{M})$ is that the condition (\ref{equ}) is fulfilled
for a density operator $\rho$ maximizing the quantity  (\ref{hbound2}).

Consider the case of overcomplete system $M_{y}=|\psi _{y}\rangle
\langle \psi _{y}|$ in $m-$
dimensional Hilbert space $\mathcal{H}$ where $\rho =I/m.$ The corresponding
ensemble is $\pi _{\rho }^{\prime}\equiv\bar{\pi}=\left\{ \frac{\langle \psi _{y}|\psi _{y}\rangle }{m},
\frac{|\psi _{y}\rangle \langle \psi _{y}|}{\langle \psi _{y}|\psi
_{y}\rangle }\right\} $ and  $\chi \left( \bar{\pi}\right) =\log m=C_{ea}(
\mathcal{M})$ (Notice that this is also equal to the classical capacity of
the c-q channel $y\longrightarrow \frac{|\psi _{y}\rangle \langle \psi _{y}|
}{\langle \psi _{y}|\psi _{y}\rangle },$ since this is the maximal possible
value). The condition (\ref{equ}) amounts to
\[
|\psi _{y}\rangle \langle \psi _{y}|\psi _{y'}\rangle \langle \psi _{y'}|
=|\psi _{y'}\rangle \langle \psi _{y'}|\psi _{y}\rangle \langle \psi _{y}|.
\]
We can always assume that the vectors $|\psi _{y}\rangle$ are all pairwise linearly independent,
then the last condition is equivalent to the fact that they form an orthonormal basis \cite{hall}.
Thus this is the only case where $C_{ea}(\mathcal{M})=C(\mathcal{M})$.

In order to pass to continuous observables we use the fact that there are
unitary operators $U_{y}$ such that $\frac{\rho ^{1/2}M_{y}\rho ^{1/2}}{
\mathrm{Tr}\rho M_{y}}=U_{y}\rho \left( y|M\right) U_{y}^{\ast }$  in the
ensemble $\pi _{\rho }^{\prime }$. Since the posterior states $\rho \left(
y|M\right) $ are well defined for arbitrary observable $M=\left\{
M(dy)\right\} $ and apriori state $\rho $ \cite{ozawa}, this opens the way
to the general definition of the ensemble $\pi _{\rho }^{\prime }=\left\{ \mathrm{Tr}
\rho M(dy),U_{y}\rho \left( y|M\right) U_{y}^{\ast }\right\} .$ Applying
this to the example 2, we find that the accessible information for the
continuous ensemble $\bar{\pi}=\left\{ \nu (d\theta ),|\theta \rangle
\langle \theta |;\theta \in \Theta \right\} $ is equal to (\ref{euler}) while $
\chi \left( \bar{\pi}\right) =\log m,$ where $\chi \left( \pi _{\rho
}^{\prime }\right) $ is defined as in (\ref{I_mod}).

The continuous ensemble is the ``Scrooge ensemble'' for the density operator $I/d$ for which the value (\ref{euler})
of the accessible information was obtained by different method in  \cite{robb}\footnote{
M. J. W. Hall, private communication.}. In \cite{asp},
Sec. IV.4 the Bayes estimation problem for this ensemble was solved; it was
shown, in particular, that with $m\to\infty$ there is no better strategy than simple guessing.
In information-theoretic scenario this would suggest zero capacity of the c-q channel
$\theta\rightarrow |\theta \rangle \langle \theta |$ which however is not the case.

In the case of infinite dimensional space with the constraint the formulas
(\ref{A1}), (\ref{A2}) should be modified as
\begin{eqnarray}
C(\mathcal{M},N) &=&\sup_{\rho :\mathrm{Tr}\rho E\leq N}A(\pi _{\rho
}^{\prime }\mathbf{),}  \label{B1} \\
C_{ea}(\mathcal{M},N) &=&\sup_{\rho :\mathrm{Tr}\rho E\leq N}\chi \left( \pi
_{\rho }^{\prime }\right) .  \label{B2}
\end{eqnarray}
Consider the canonical observable (\ref{canobs}) with the state $\rho _{N}.$
The corresponding ensemble is
\[
\bar{\pi}=\left\{ \langle z|\rho _{N}|z\rangle \frac{d^{2}z}{\pi },\frac{
\rho _{N}^{1/2}|z\rangle \langle z|\rho _{N}^{1/2}}{\langle z|\rho
_{N}|z\rangle };\,z\in\mathbb{C}\right\} .
\]
By making computation in the Fock basis, we have $\langle z|\rho _{N}|z\rangle =(N+1)^{-1}\exp \left( -\frac{|z|^{2}}{
N+1}\right) $ and $\rho _{N}^{1/2}|z\rangle =c\left\vert \sqrt{\frac{N}{N+1}}
z\right\rangle ,$ so that the ensemble states are $\left\vert \sqrt{\frac{N}{
N+1}}z\right\rangle \left\langle \sqrt{\frac{N}{N+1}}z\right\vert .$ But
with the change of variable $z^{\prime }=\sqrt{\frac{N}{N+1}}z$ this
ensemble is equivalent to the ensemble
\[
\bar{\pi}^{\prime }=\left\{ \exp \left( -\frac{|z|^{2}}{N}\right) \frac{
d^{2}z}{\pi N},|z\rangle \langle z|\right\} .
\]
From (\ref{B1}), (\ref{B2}) it follows
\begin{eqnarray*}
A(\bar{\pi}^{\prime }\mathbf{)} &\mathbf{=}&C(\mathcal{M},N)=\log (N+1), \\
\chi (\bar{\pi}^{\prime }\mathbf{)} &\mathbf{=}&C_{ea}(\mathcal{M},N)=\log
(N+1)+\log \left( 1+\frac{1}{N}\right) ^{N}.
\end{eqnarray*}
The last expression is also equal to the constrained classical capacity of
the c-q channel $z\longrightarrow |z\rangle \langle z|$.

\section{Appendix}

1. In the paper \cite{bsst} it was shown that the depolarizing
channel
\begin{equation}
\Phi \left( \rho \right) =(1-p)\rho +p\frac{I}{m}  \label{dep}
\end{equation}
in $m$ dimensions is entanglement-breaking for $p\geq \frac{m}{m+1}$ : ``The
simulation is performed by having Alice measure in a pre-agreed random
basis, send Bob the result through a $m$-ary symmetric noisy classical
channel, after which he re-prepares an output state in the same basis''. We
will supply an analytical proof by showing that

\bigskip
\[
\frac{1}{m+1}\rho +\frac{m}{m+1}\frac{I}{m}=m\int_{\Theta }|\theta \rangle
\langle \theta |\rho |\theta \rangle \langle \theta |\nu (d\theta )
\]
for all density operators $\rho ,$ which means that the depolarizing channel
with $p=\frac{m}{m+1}$ is entanglement-breaking \cite{ebr}. Then the depolarizing channel for $p>\frac{m}{m+1}$ can be represented
as mixture of this channel and completely depolarizing channel $\rho
\rightarrow \frac{I}{m},$ which are both entanglement-breaking.

It is sufficient to establish (\ref{dep}) for all $\rho =|\theta ^{\prime
}\rangle \langle \theta ^{\prime }|,$ $\theta ^{\prime }\in \Theta .$
Consider the operator
\[
\sigma =m\int_{\Theta }|\theta \rangle \langle \theta |\theta ^{\prime
}\rangle \langle \theta ^{\prime }|\theta \rangle \langle \theta |\nu
(d\theta ).
\]
It has trace 1 since
\[
\mathrm{Tr}\sigma =m\int_{\Theta }|\langle \theta |\theta ^{\prime }\rangle
|^{2}\nu (d\theta )=-m\int\limits_{0}^{1}r^{2}d(1-r^{2})^{m-1}=1
\]
by (\ref{formula}). From this (\ref{resid}) follows by polarization.

Next, $\sigma $ commutes with all the unitaries leaving invariant $|\theta
^{\prime }\rangle ,$ hence it has the form
\[
\left( 1-p\right) |\theta ^{\prime }\rangle \langle \theta ^{\prime }|+p
\frac{I}{m}.
\]
To find $p$ take $\langle \theta ^{\prime }|\sigma |\theta ^{\prime }\rangle
,$ then we obtain
\[
m\int_{\Theta }|\langle \theta |\theta ^{\prime }\rangle |^{4}\nu (d\theta
)=-m\int\limits_{0}^{1}r^{4}d(1-r^{2})^{m-1}=\left( 1-p\right) +\frac{p}{m}.
\]
Computing the integral with the formula (\ref{formula}) we get the value $
\frac{2}{m+1},$ whence $p=\frac{m}{m+1}.$

2. Proof of the formula
\[
\int\limits_{0}^{1}m(1-u)\ln m(1-u)du^{m-1}=\ln m-\sum_{k=2}^{m}\frac{1}{k}
.
\]
Splitting the integral and integrating by parts we obtain
\begin{eqnarray*}
&&m\ln m\int\limits_{0}^{1}(1-u)du^{m-1}+ \\
&+&m\int\limits_{0}^{1}(1-u)\ln (1-u)du^{m-1} \\
&=&\ln m+m\int\limits_{0}^{1}u^{m-1}\left[ 1+\ln (1-u)\right] du \\
&=&\ln m+1+\int\limits_{0}^{1}\ln (1-u)d\left( u^{m}-1\right) \\
&=&\ln m+1-\int\limits_{0}^{1}\frac{\left( u^{m}-1\right) }{u-1}du \\
&=&\ln m+1-\int\limits_{0}^{1}\sum_{k=0}^{m-1}u^{k}du.
\end{eqnarray*}
\bigskip

\textbf{Acknowledgement}. The author thanks P.W. Shor, M.E. Shirokov and M.
J. W. Hall for enlightening comments.

\end{document}